\documentclass{aa}

\pdfoutput=1

\usepackage[varg]{txfonts}
\usepackage{enumerate}
\usepackage{enumitem}
\usepackage{float}
\usepackage{amsmath}

\usepackage{natbib}
\bibpunct{(}{)}{;}{a}{}{,} 

\newcommand{\fref}[1]{Fig. \ref{#1}}			
\newcommand{\aref}[1]{Appendix \ref{#1}}		

\newcommand{\eref}[1]{Eq. \eqref{#1}}			
\newcommand{\erefnp}[1]{Eq. \ref{#1}}			
\newcommand{\erefp}[1]{(Eq. \ref{#1})}			
\newcommand{\erefCon}[2]{Eqs. \eqref{#1}--\eqref{#2}}	
\let\d=\dsub                 
\newcommand{\d}{\textnormal{d}}
\let\t=\tsub                 
\newcommand{\t}[1]{\text{\textnormal{#1}}}  

\begin{document}

\title{Formation of pebble-pile planetesimals}

\subtitle{}

\author{Karl Wahlberg Jansson
\and Anders Johansen}

\institute{Lund Observatory, Department of Astronomy and Theoretical Physics, Lund University, Box 43, SE-221 00 Lund, Sweden \newline\email{kalle@astro.lu.se}}

\date{Recieved ... / Accepted ... }

\abstract{Asteroids and Kuiper belt objects are remnant planetesimals from the epoch of planet formation. The first stage of planet formation is the accumulation of dust and ice grains into mm-cm-sized pebbles. These pebbles can clump together through the streaming instability and form gravitationally bound pebble `clouds'. Pebbles inside such a cloud will undergo mutual collisions, dissipating energy into heat. As the cloud loses energy, it gradually contracts towards solid density. We model this process and investigate two important properties of the collapse: (i) the timescale of the collapse and (ii) the temporal evolution of the pebble size distribution. Our numerical model of the pebble cloud is zero-dimensional and treats collisions with a statistical method. We find that planetesimals with radii larger than $\sim$100 km collapse on the free-fall timescale of about 25 years. Lower-mass clouds have longer pebble collision timescales and collapse much more slowly, with collapse times of a few hundred years for 10-km-scale planetesimals and a few thousand years for 1-km-scale planetesimals. The mass of the pebble cloud also determines the interior structure of the resulting planetesimal. The pebble collision speeds in low-mass clouds are below the threshold for fragmentation, forming pebble-pile planetesimals consisting of the primordial pebbles from the protoplanetary disk. Planetesimals above 100 km in radius, on the other hand, consist of mixtures of dust (pebble fragments) and pebbles which have undergone substantial collisions with dust and other pebbles. The Rosetta mission to the comet 67P/Churyumov-Gerasimenko and the New Horizons mission to Pluto will both provide valuable information about the structure of planetesimals in the Solar System. Our model predicts that 67P is a pebble-pile planetesimal consisting of primordial pebbles from the Solar Nebula, while the pebbles in the cloud which contracted to form Pluto must have been ground down substantially during the collapse.}


\maketitle

\section{Introduction}\label{sec:intro}

In the process of forming planets the initially $\mu$m-sized dust and ice particles in the primordial protoplanetary disk grow to planets of sizes up to $\sim$10$^4$ km. This is believed to occur in distinct stages where different physical growth processes dominate \citep{safronov69}. The growth from dust to planetesimals occurs at first via direct sticking but is later aided by the increase in the local particle density by concentration in the turbulent gas \citep{johansen14}. Small particles can stick together via surface forces, but when they reach sizes of mm-cm growth stops as collisions lead to bouncing \citep{zsom10} and fragmentation \citep{brauer08, birnstiel09}. The two leading theories for the growth mechanism beyond this point are: 1) hierarchical coagulation and 2) gravitational instability. 

Growth by coagulation can proceed via mass transfer as small impactors deposit part of their mass when hitting a large target particle \citep{wurm05}. The resulting growth is nevertheless slow, resulting in the formation of 100 m-scale planetesimals in the asteroid belt in 1 Myr \citep{windmark12}; and growth rates would be even lower in the Kuiper belt.



Planetesimal formation via gravitational instability is not without problems either. Gravitational instability of a region in an unperturbed protoplanetary disk can be easily stopped by Kelvin-Helmholz turbulence in the mid-plane layer \citep{goldreich73, weidenschilling80}. However, the presence of gas in the disk can cause particles to clump together through the streaming instability \citep{youdin05} and lead to high particle overdensities. Simulations of this mechanism in protoplanetary disks result in the formation of gravitationally bound pebble clouds with masses comparable to $\sim$100 km solid planetesimals or larger \citep[e.g.][]{johansen09, johansen12}.

An advantage with gravitational instability is that, if the internal angular momentum of a gravitationally unstable pebble cloud is large, it naturally forms a binary planetesimal with mass ratio of $\sim$1 \citep{nesvorny10}. The components form from the same material, so they will also have the same chemical composition and colour, in agreement with what is observed for binary Kuiper belt objects \citep{benecchi09}. The Kuiper belt as a whole has a wide colour distribution, so this suggests that the components in binary Kuiper belt objects formed together; a similar co-formation explanation is not possible in models of binary formation through three-body encounters \citep{goldreich02} in the hierarchical coagulation formation mechanism. 

\citet{nesvorny10} modelled the particle cloud with an N-body method with perfect sticking. The goal of this paper is to use more realistic collision outcomes based on laboratory experiments \citep{guttler10} to model the collapse. Inelastic collisions lead to energy dissipation and contraction of the cloud, which in turn releases gravitational energy into random pebble motion. As a pebble cloud undergoes collapse, the relative speed between the pebbles can, depending on the mass of the cloud, reach values large enough to result in fragmentation. Low-mass clouds experience no fragmentation and the planetesimal will be formed from primordial pebbles. The pebbles in high-mass clouds, on the other hand, are ground down to dust during the collapse. Depending on the degree of fragmentation, the porosity and density of the resulting planetesimal will vary. In the case of Kuiper belt objects, measured densities range from below 1 g cm$^{-3}$ for smaller bodies (radii $\lesssim 200$ km) up to $\sim$2.5 g cm$^{-3}$ for Eris (radius $\sim$1000 km) \citep[Fig. 3 in][]{brown13}. The density of a Kuiper belt object is not only determined by the porosity but also by the composition. A large ice fraction results in a lower density, suggesting that more massive objects are both less porous and have a higher rock fraction. 

Observations of comet nuclei in the solar system indicate densities of $\sim$0.5 g cm$^{-3}$ \citep{weissman04,davidsson07} due to high porosity combined with large ice fraction. High porosities can be either meso-scale (rubble piles) or small scale (pebble piles). A pebble pile could originate from the collapse of a low-mass pebble cloud where the collision speeds never reach the fragmentation limit. Studies of the tidal break-up of the comet Shoemaker-Levy 9 showed that its internal strength was as low as expected for a rubble (or pebble) pile \citep{asphaug96}. Observations of the comet 67P/Churyumov-Gerasimenko \citep{bauer12} show that mm-sized pebbles are released from the surface of the comet. \citet{skorov12} model this pebble release process with a two-layered comet nucleus consisting of an ice-free surface layer residing on top of a mixture of refractory pebbles and ice aggregates. These observations agree with the idea of comets being piles of mm-cm-sized pebbles. The interior structure of comets will be further investigated by the Rosetta mission which will put a lander on 67P/Churyumov-Gerasimenko in November 2014.

An additional motivation for this paper is the NASA mission New Horizons which is on its way to the outer regions of our Solar System. It is scheduled to arrive at the Pluto-Charon system in July 2015 \citep{stern08}. After a fly-by it is planned to encounter and investigate one or two additional Kuiper belt objects. The New Horizons probe will measure the mass, binarity, surface composition and temperature in such fly-bys. As no currently known Kuiper belt objects are within the range of New Horizons after the fly-by of Pluto a survey is ongoing to detect possible candidates \citep{buie12}.

We have developed a model for simulating the collapse of gravitationally bound pebble clouds. With this model we investigate e.g.\ the collapse time and particle size distribution as function of planetesimal size. The paper is organized as follows. We present our zero-dimensional model of the cloud, the model of collisional outcomes and the numerical scheme in Section \ref{sec:model}. The simulations of clouds with different masses and initial pebble sizes are presented in Section \ref{sec:sims} and the results are discussed in Section \ref{sec:conc}. In Appendix \ref{app:A} we present an analytic description of the collapse of a low-mass pebble cloud experiencing only bouncing collisions.

\section{Model}\label{sec:model}

We start with a gravitationally bound cloud of pebbles formed for example through the streaming instability. As the pebbles move around they dissipate energy in inelastic collisions with each other leading to a contraction of the cloud. Since gravitationally bound systems have a negative heat capacity, the pebbles will increase their speed when the cloud loses energy. This together with the increased density as the cloud contracts will lead to higher collision rates and faster energy dissipation. The result is a runaway collapse. With some approximations (e.g.\ only bouncing collisions, equal particle size and immediate virialization) this collapse can be derived analytically (See \aref{app:A}). Including more realistic collision physics requires numerical solution.

\subsection{Parametrisation of the cloud size and energy}

Our model is zero-dimensional, so it requires that the cloud is completely uniform and that there is no net rotation. This is of course not completely physical but it is a first step towards a more complete model. We also assume that the cloud is in, or strives to get into, virial equilibrium. The cloud is initially set in virial equilibrium so that the initial total energy, $E_0$, is

\begin{align}
 E_0=T_0+U_0&=-T_0=-\frac{1}{2}\sum_{i=1}^Nm_iv_i^2=-\frac{1}{2}Mv_\t{vir,0}^2  \nonumber \\
            &=\frac{1}{2}U_0=-\frac{3}{10}\frac{GM^2}{R_0}\ . \label{eq:virialEnergy}
\end{align}

\noindent Here $T_0$ is the initial total kinetic energy (the sum of the kinetic energy of each individual particle with mass $m_i$ and speed $v_i$), $M$ is the total mass of the pebble cloud, $v_\t{vir,0}$ is the virial speed calculated from the initial kinetic energy, $U_0$ is the initial potential energy, $G$ is the gravitational constant and $R_0$ is the initial radius of the cloud. During the collapse we can follow the cloud's evolution through three parameters: $\eta_\t{eq}$, $\eta$ and $\eta_\t{K}$. The parameter $\eta_\t{eq}$ is defined as the initial energy divided by the current total energy $E(t)$,

\begin{align}
 \eta_\t{eq}(t)\equiv \frac{E_0}{E(t)}\ .
\end{align}

The parameter $\eta$ is defined as the current size, $R(t)$, of the cloud divided by the initial size, 

\begin{align}
 \eta(t)\equiv \frac{R(t)}{R_0}=\frac{U_0}{U(t)}\ . \label{eq:eta}
\end{align}

Finally $\eta_\t{K}$, which we use to generate relative speeds when the cloud is not in virial equilibrium, is defined as the initial kinetic energy divided by the current kinetic energy, $T(t)$,

\begin{align}
 \eta_\t{K}(t)\equiv \frac{T_0}{T(t)}\ . \label{eq:etaK}
\end{align}

The cloud is bound and contracting, so $\eta_\t{eq}$ and $\eta$ have values between 0 and 1, while $\eta_\t{K}$ can have any value since kinetic energy is both dissipated in collisions and released in the gravitational collapse. These three parameters do not need to have the same value since the cloud does not virialize immediately after a collision. We assume that the cloud virializes on a timescale of the free-fall time of the cloud. At a certain size of the cloud, $\eta$, we can get the maximum possible change in $\eta$, $\delta\eta_\t{max}$, by assuming that the cloud has been free-falling from $R_0$. This gives

\begin{figure*}[t!]
 \begin{center}
  \resizebox{9cm}{!}{\includegraphics{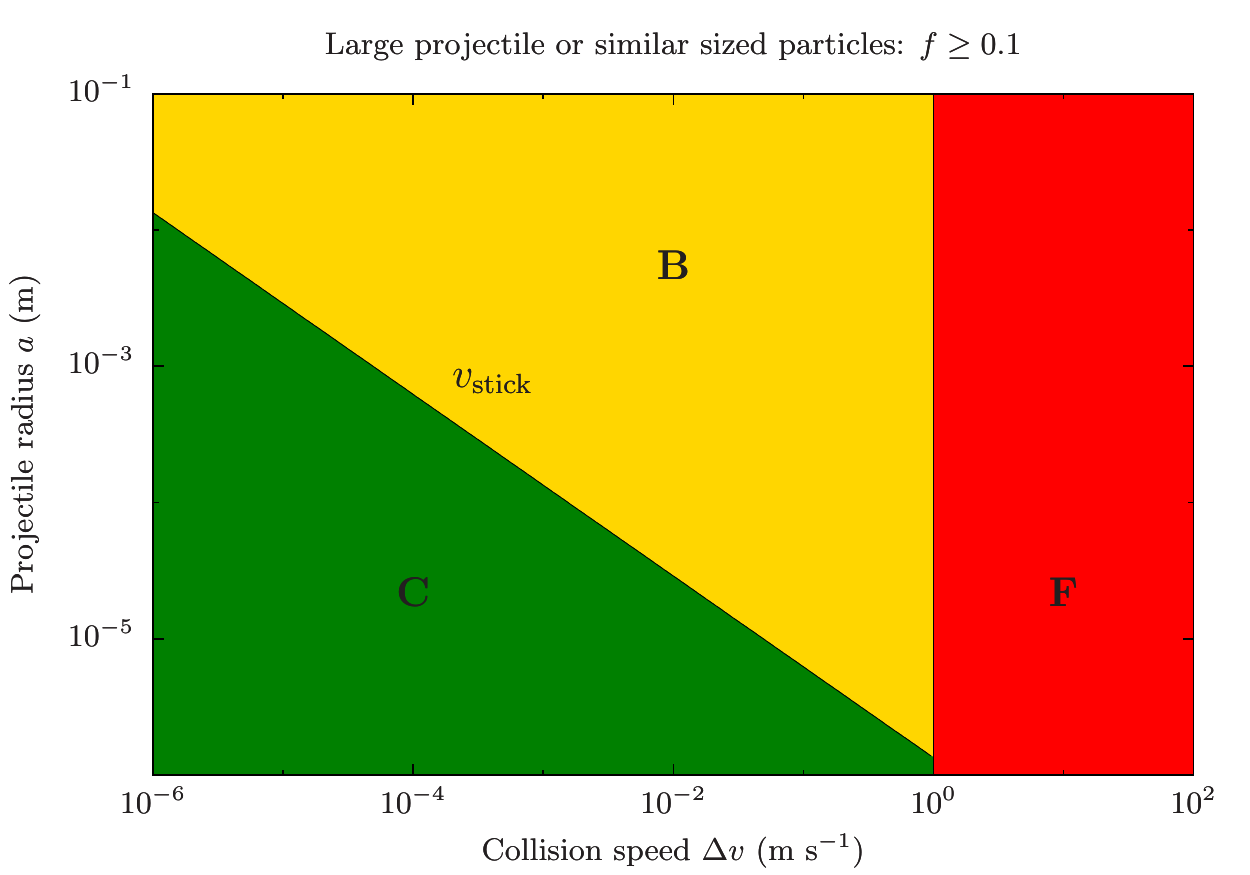}}
  \resizebox{9cm}{!}{\includegraphics{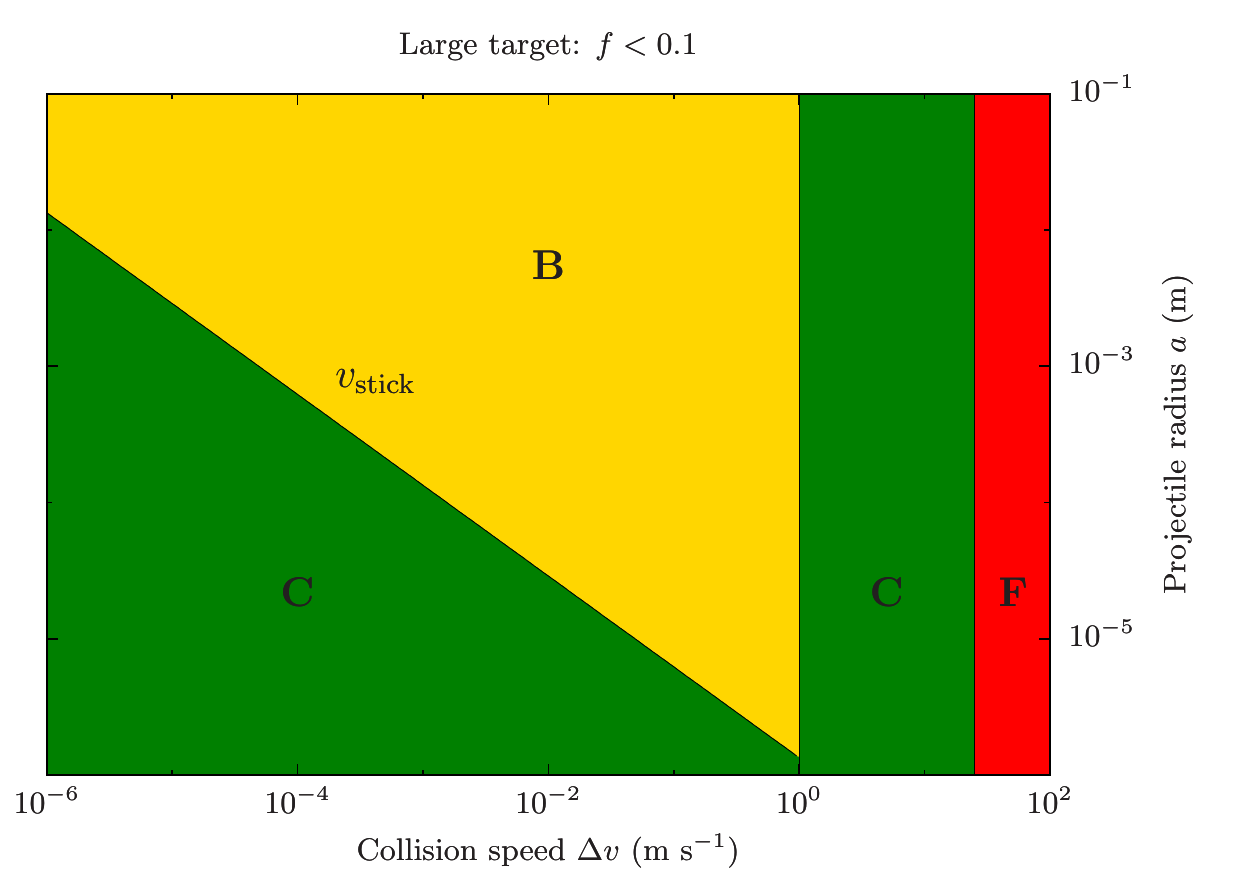}}
  \caption{Collisional outcomes as a function of projectile radius, $a$, collision speed, $\Delta v$, and relative particle size, $f\equiv m_2/m_1$. Green regions indicate coagulation, yellow regions bouncing and red regions fragmentation. Outcome regions taken from \citet{guttler10}.} \label{fig:outcomeFig}
 \end{center}
\end{figure*}

\begin{align}
 \delta\eta_\t{max}&=\frac{\delta R_\t{max}}{R_0}=\frac{1}{R_0}\left.\frac{\d R}{\d t}\right\vert_\t{ff}\delta t=-\frac{1}{R_0}\sqrt{\frac{2GM}{R_0}}\sqrt{\frac{R_0-R}{R}}\delta t \nonumber \\
 &=-\frac{\pi}{2}\frac{1}{t_\t{ff,0}}\sqrt{\frac{1-\eta}{\eta}}\delta t\ , \label{eq:deltaEtaMax}
\end{align}

\noindent where $t_\t{ff,0}$ is the initial free-fall time. Here we have used the free-fall speed at $R$ starting from $R_0$,

\begin{align}
 v_\t{ff}=\sqrt{\frac{2GM}{R_0}}\sqrt{\frac{R_0-R}{R}}\ .
\end{align}

The `desired' change in $\eta$ after a dissipative collision is $\delta\eta=\eta_\t{eq}-\eta$. If $\vert\delta\eta\vert<\vert\delta\eta_\t{max}\vert$ the new value of $\eta$ is equal to the equilibrium value, otherwise the new value is the old plus $\delta\eta_\t{max}$, since there is not enough time for the system to virialize in the time step $\delta t$.

When it comes to updating $\eta_\t{K}$ we need to take into account that kinetic energy is both dissipated in the collisions and released when the cloud contracts, so using Eqs. \eqref{eq:virialEnergy}, \eqref{eq:eta} and \eqref{eq:etaK} we obtain

\begin{align}
 \delta\eta_\t{K}&=-\frac{T_0}{T^2}\delta T=-\eta_\t{K}^2\frac{\delta T}{T_0}=-\eta_\t{K}^2\frac{\delta E-\delta U}{T_0} \nonumber \\
 &=\eta_\t{K}^2\left(\underbrace{\frac{\delta E}{E_0}}_{>0}+\underbrace{\frac{2}{\eta^2}\delta\eta}_{<0}\right)\ , \label{eq:deltaEtaT}
\end{align}

\noindent where $\delta E$ is the energy lost in the collision (see Section \ref{sec:collOut}), $\delta U$ is the gravitational energy released and $\delta\eta$ is the previously calculated change in $\eta$ (could be $\delta\eta_\t{max}$). The first term in \eref{eq:deltaEtaT} decreases the kinetic energy (energy dissipation in collisions) while the second term increases the kinetic energy (release of gravitational potential in the collapse). If the virialization is too slow (i.e. if the next collision occurs before the gravitational energy has been released), $\delta\eta_\t{K}>\delta\eta$ and the particles will achieve sub-virial velocities.

\subsection{Collisional outcomes} \label{sec:collOut}

For the cloud to collapse it needs to dissipate a lot of kinetic energy. In our model this is done by inelastic collisions between the particles. We use the tabulated laboratory measurements of \citet{guttler10} to get the outcome of a collision between a target particle (subscript 1) and a projectile (subscript 2). The outcome depends on the relative speed, $\Delta v$, and relative size, $f\equiv m_2/m_1$, of the particles. The collisions are split into two groups:

\begin{enumerate}[label=\roman*)]

 \item Two similar-sized particles colliding or a larger particle, $2$, colliding with a small particle, $1$: $f\geq 0.1$

 \item A small particle, $2$, `running' into (colliding with) a larger particle, $1$: $f<0.1$

\end{enumerate} 

We do a simple interpretation of Fig. 11 in \citet{guttler10}, for collisions between compact particles:

\medskip
\begin{center}
\begin{tabular}{c|c|c}
 $\Delta v$ (m s$^{-1}$) & $f\geq 0.1$ & $f<0.1$ \\ \hline
 $\Delta v< v_\t{stick}$ & C & C \\ \hline
 $v_\t{stick}\leq\Delta v< 1$ & B & B \\ \hline
 $1\leq\Delta v< 25$ & F & C \\ \hline
 $\Delta v \geq 25$ & F & F  
\end{tabular}
\end{center}

\noindent Here C denotes coagulation, B bouncing and F fragmentation. The threshold speed for sticking, $v_\t{stick}$, is taken from \cite{guttler10}, as

\begin{align}
 v_\t{stick} = \sqrt{\frac{5\pi a_0F_\t{roll}}{\mu}}\ ,
\end{align}

\noindent where $\mu=m_1m_2/(m_1+m_2)$ is the reduced mass, $a_0$ is the monomer radius and $F_\t{roll}$ is the rolling force of the monomers\footnote{We use $F_\t{roll}=(8.5\pm 1.6)\times 10^{-10}$ N for SiO$_2$ spheres with $a_0\sim 1$ $\mu$m from \citet{heim99}.}. The coagulation outcome of collisions with $f<0.1$ and $1$ m s$^{-1}$ $\leq\Delta v< 25$ m s$^{-1}$ is explained by mass transfer or penetration \citep{wurm05}. A map of collision outcomes in $(\Delta v,a)$-space is shown in \fref{fig:outcomeFig}.

Next we determine the amount of energy dissipated, $\delta E$, in a collision between target particle $1$ and projectile $2$. In the case of bouncing we use conservation of momentum and the coefficient of restitution, $C_\t{R}$, to get the change in energy, 

\begin{align}
 \delta E=-\frac{1}{2}\mu\Delta v^2\left(1-C_\t{R}^2\right)\ . \label{eq:deltaE}
\end{align}

For simplicity we assume that all kinetic energy in the direction of relative motion is dissipated, $C_\t{R}=0$. Particle properties like mass and radius remain constant in a bouncing collision. In the case of coagulation we use the same equation but the difference from bouncing is that the mass of the target increases to $m_1' = m_1+m_2$ and hence the radius also changes.

A constant $C_R=0$ is not physical since the collisions might not be completely inelastic. \citet{higa96} investigate the coefficient of restitution for collisions between ice pebbles. They find that $C_R$ depends on the collision speed and for low values ($\Delta v\lesssim 0.35$ m s$^{-1}$) the collisions are almost elastic ($C_R$$\sim$$0.9$). For higher velocities, on the other hand, $C_R$ decreases with increasing collision speed and reaches values close to zero at a few m s$^{-1}$. Compared to our simulations these experiments investigate ice pebbles while we consider silicates. A non-zero coefficient of restitution would decrease the amount of energy dissipated in each bouncing collision which, in turn, would increase the timescales of the collapses. Another implication is that the clouds would get more time to virialize and the collision speeds becomes a bit higher. However, the value of the coefficient of restitution would need to be relatively high to give a significant effect to the collapse since it is the square of $C_R$ that appears in \eref{eq:deltaE}.

The effect of fragmentation is a bit different. In a collision between two particles with mass $m_1$ and $m_2$ we have, in the center-of-mass frame, an available collision energy of

\begin{align}
 E_\t{coll}=\frac{1}{2}\mu\Delta v^2\ . \label{eq:E_coll}
\end{align}

\noindent Next we assume that all particles are composed of $\mu$m-sized monomers of mass $m_0$. The energy with which any contact surface of two monomers is held together is \citep{dominik97,blum00}

\begin{align}
 E_\t{roll}=\frac{1}{2}\pi a_0F_\t{roll}\ . \label{eq:E_roll}
\end{align}

With \erefCon{eq:E_coll}{eq:E_roll} we get the number of bonds that can be broken by the collision as $N_\t{cs}$$\sim$$E_\t{coll}/E_\t{roll}$. For a compact aggregate of monomers, each monomer has, on average, three contact surfaces \citep{zsom08} so the energy required to completely fragment a particle is

\begin{align}
 E_\t{frag} \sim 3\frac{m_1}{m_0}E_\t{roll}\ .
\end{align}

\begin{figure*}
 \begin{center}
  \resizebox{15cm}{!}{\includegraphics{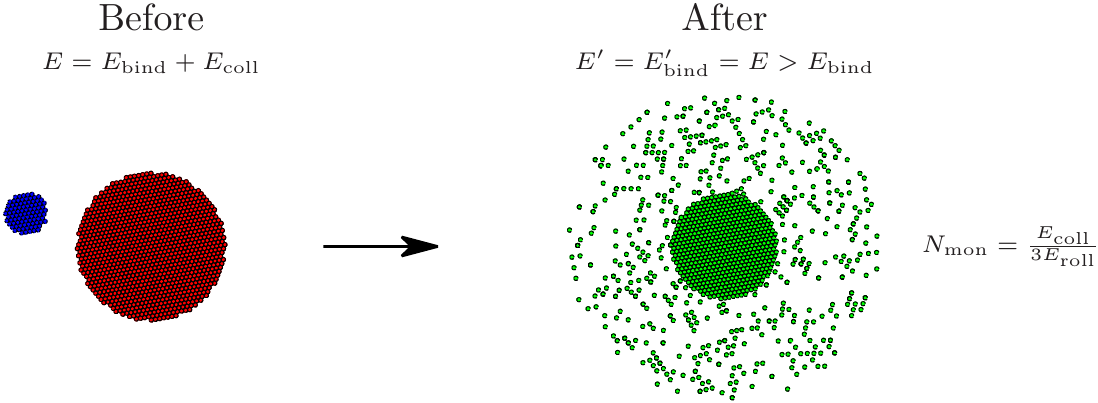}}
  \caption{In our model pebbles are composed of $\mu$m-sized monomer particles. In collisions resulting in partial fragmentation monomers are removed one by one to simulate erosion. The binding energy of the fragment, $E_\t{bind}'$, is larger (closer to zero) than the binding energy of the original particle, $E_\t{bind}$, since the fragment consists of fewer monomers. The number of monomers released, $N_\t{mon}$, is equal the amount of contact surfaces the collision energy can break divided by three contact surfaces per monomer, $N_\t{mon}=E_\t{coll}/\left(3E_\t{roll}\right)$.} \label{fig:partialFragFig}
 \end{center}
\end{figure*}

If $E_\t{coll}\geq E_\t{frag}$ the whole particle is destroyed and $\mu$m-sized monomers are produced. The energy dissipated is $\delta E=-E_\t{frag}$, as not all collision energy is needed to fragment the target. If, however, $E_\t{coll}<E_\t{frag}$ some part of the old particle survives with $N'$ monomers and $\sim$$3N'$ contact surfaces. For simplicity we assume that in the collision the monomers are removed one by one (see sketch in \fref{fig:partialFragFig}). The mass of the fragment becomes

\begin{align}
 m_1'=m_1-\frac{E_\t{coll}}{3E_\t{roll}}m_0\ . \label{eq:newMass}
\end{align}

\noindent In this case the energy dissipated is $\delta E=-E_\t{coll}$, all the collision energy goes into partially fragmenting the target. 

To make the simulations more physically realistic we add an impact parameter, $b$, to all collisions. The value of $b$ is randomly picked between 0 and the sum of the radii ($a_1$ and $a_2$) of the two particles with higher probability for larger impact parameters (larger area),

\begin{align}
 \d P(b)=\frac{2b}{\left(a_1+a_2\right)^2}\d b\ .
\end{align}

The effect of a non-zero impact parameter is that the effective relative speed decreases. This means that the outcome of the collision can change and the energy dissipated will be lower.

\subsection{Representative particle approach} \label{sec:zsom}

To model the collapse of a planetesimal mass cloud of mm-cm-sized pebbles, formed e.g.\ through the streaming instability  \citep{johansen09, johansen12}, it is not possible to follow the evolution of every single particle. A Pluto-mass planetesimal split into cm-sized pebbles results in $N$$\sim$$10^{24}$ particles. Therefore we use the representative particle approach of \citet{zsom08} where the particles are grouped into a smaller number (manageable by a computer) of superparticles, each representing a large number of particles with identical properties. We repeat the main steps of the \citet{zsom08} algorithm here.

\medskip
The representative particle approach is a Monte Carlo model that uses the collision rates of all possible collision pairs to find which particles collide and the times between collisions. Initially a large number, $N$, of physical particles are defined. From these we randomly select a smaller number, most simulations are done with $n= 1000$, representative particles which we follow the evolution of. Each representative particle, $i$, has its own properties like mass or velocity. One can think of a representative particle as one member of a swarm of identical physical particles: if the properties of the representative particle changes the properties of all the particles within the swarm changes. In the simulation the total mass of each of these swarms remains constant ($M_i=M/n$) so the number of physical particles in a swarm will change.

\medskip
To follow the evolution of the cloud we need to know with what rate a representative particle collides with any other particle. Assume that $n$ is large enough so the distributions of the properties of the representative particles is a good representation of the true distributions of all $N$ physical particles. We then get the collision rate between representative particle $i$ and a physical particle with the same properties as representative particle $k$ to be

\begin{align}
 r_i(k)=n_k\sigma_{ik}\Delta v_{ik}\ , \label{eq:rik}
\end{align}

\noindent where $n_k$ is the number density of physical particles represented by representative particle $k$, $\sigma_{ik}$ is the cross-section between two such particles and $\Delta v_{ik}$ is the relative speed. In our simulations we generate the relative speeds from the kinetic energy of the cloud, $\eta_\t{K}$, and assuming a maxwellian distribution of the relative speeds,

\begin{align}
\t{d}P(\Delta v)=\frac{1}{2\sqrt{\pi}}\frac{\Delta v^2}{\sigma^3}e^{-\Delta v^2/4\sigma^2}\t{d}(\Delta v)\ , \label{eq:dv_prob}
\end{align}

\noindent where $\sigma=v_\t{vir,0}/\sqrt{3\eta_\t{K}}$ is the one-dimensional velocity dispersion in the cloud and we have included the fact that the relative speeds are, on average, a factor $\sqrt{2}$ larger than the individual speeds. For strongly dissipative collisions we do not expect energy equipartition to occur and hence the small particles should have similar velocities as the large particles. From \eref{eq:rik} we calculate the rate with which representative particle $i$ collides with any particle (it can collide with particles in its own swarm as well),

\begin{align}
 r_i=\sum_{k=1}^n r_i(k)\ ,
\end{align}

\noindent and the total rate of collisions for any of the representative particles,

\begin{align}
 r=\sum_{i=1}^n r_i\ . \label{eq:rate}
\end{align}

\noindent From this total rate and the assumption that the collision process is a Poisson process we get the exponentially distributed time until next collision,

\begin{align}
 \delta t=-\frac{1}{r}\log(\t{rnd(seed}))\ ,
\end{align}

\noindent where $\t{rnd(seed)}$ is a random number uniformly distributed between 0 and 1. Next we choose which particles collide with two more random numbers. First the representative particle $i$ with the probability

\begin{align}
 P(i)=\frac{r_i}{r}\ ,
\end{align}

\noindent and then the physical particle $k$, given representative particle $i$, with the probability

\begin{align}
 P(k\vert i)=\frac{r_i(k)}{r_i}\ .
\end{align}

Knowing which two particles collide we calculate the outcome and effects of such a collision. For this we use the equations in Section \ref{sec:collOut} changing the subscripts 1 and 2 to $i$ and $k$, i.e. the representative particle is the target and the physical particle is the projectile. We also need to take into consideration that all equations in Section \ref{sec:collOut} are for one collision only. Therefore, when calculating the energy dissipated we need to multiply the equations with the number of physical particles in the swarm, $N_i=\frac{M_i}{m_i}$. Another difference arises when we pick the new mass of the representative particle after a fragmenting collision. If we have complete fragmentation the new mass is simply the monomer mass $m_0$. If, on the other hand, we have partial fragmentation, we pick the new mass from a mass-weighted bimodal distribution defined by the value $m_i'$ (\eref{eq:newMass}). We pick a random number uniformly distributed between 0 and $m_i$, if it is $\leq m_i'$ the new mass is $m_i'$ otherwise the new mass is $m_0$. Basically we randomly select a monomer in the original particle and see where it ends up.

After a collision, say representative particle $j$ collided with physical particle $l$, the properties of the representative particle has changed and the rate matrix, \eref{eq:rik}, needs to be updated. Representative particle $l$ has not been involved in the collision and since we are only looking at the evolution of the representative particles the only elements that needs to updated are the ones with representative particle $i=j$ and physical particle $k=j$ [one row and one column in $r_i(k)$]. The symmetry is restored when representative particle $l$ collides with physical particle $j$. Here we also take advantage of the $\eta$-parametrization of the cloud. After every collision the size of the cloud and the virial speed changes which means that the number density, $n_i$, and relative speed, $\Delta v_{ik}$, changes for all particles and the full rate matrix should need an update. But since $n_i\propto R(t)^{-3}\propto \eta^{-3}$ and $\Delta v_{ik}\propto T(t)^{1/2}\propto\eta_\t{K}^{-1/2}$ by keeping track of the $\eta$s it is not necessary to update all $r_i(k)$-elements every time step. Other quantities, like $r$ \erefp{eq:rate}, have simple scalings with $\eta$ and $\eta_\t{K}$.

Finally we also need to update the cumulative distribution functions (CDFs) for picking the representative and physical particles. The CDFs for the physical particles with $i\neq j$ only need to be updated from element $j$ and onwards since the elements $r_i(k)$ with $k\neq j$ are unchanged. The CDF for the representative particle and the CDF for physical particle with $i=j$ need a complete update. The result of this is that the time for updating scales as $\sim n\times(n-j+1)$.

\begin{figure*}
 \begin{center}
  \resizebox{15cm}{!}{\includegraphics{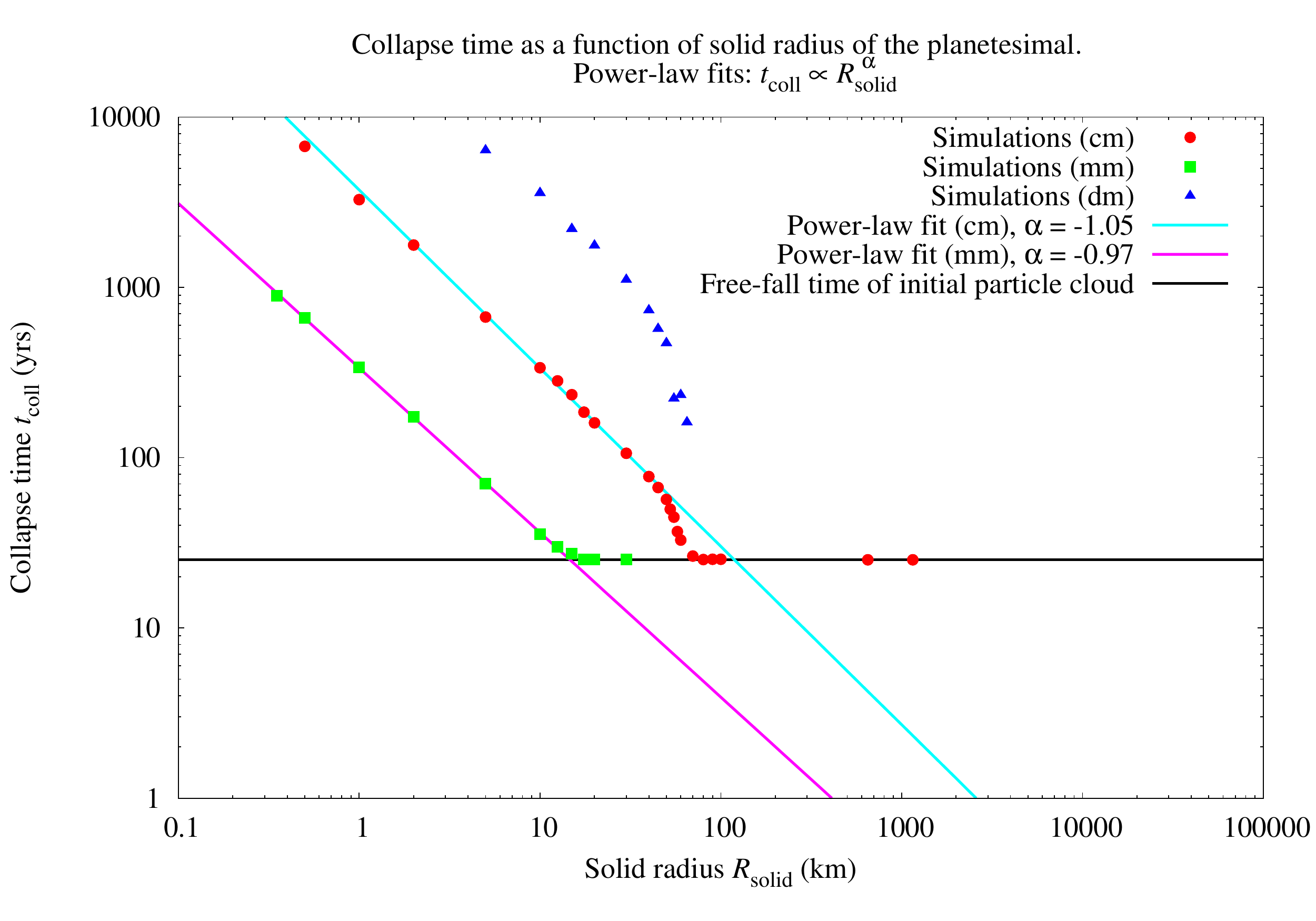}}
  \caption{Collapse time as a function of solid planetesimal radius for clouds with initially monodisperse mm-, cm- or dm-sized pebbles. Power-law fits to the data and the free-fall time of the initial cloud ($t_\t{ff}\sim 25.1$ yrs) are added to the plot. For small planetesimals ($\lesssim 50$ km) the collapse time is inversely proportional to the planetesimal radius and increases linearly with the pebble size. This is the same relation as we find in \aref{app:A} where we assume bouncing as the only collisional outcome. For larger planetesimals ($R_\t{solid}\gtrsim 50$ km) the collapse time drops rapidly as pebbles fragment and the number density and collision rates increases. The collapse time is, however, limited downwards by free-fall, so for even larger planetesimals with $R_\t{solid}\gtrsim70$ km the collapse time is equal to the free-fall time of the initial cloud.} \label{fig:tCollFig}
 \end{center}
\end{figure*}

\subsection{Pooling of collisions}

Due to high relative speeds some collisions in the simulations can lead to complete fragmentation. The collision rate between a large particle and monomers increases drastically because of the large number density of small particles (see \erefnp{eq:rik}, $n_k\propto m_k^{-1}$). This, in turn, leads to very short time steps and many collisions with almost no impact on the system. We solve this issue by pooling up collisions if the mass ratio of the two particles is too small, $f=m_k/m_i<f_\t{crit}$. We pool collisions using the method suggested in \citet{zsom08}: we lower the probability for collisions with small mass ratios (projectile mass $\ll$ target mass), but increase the number of physical collisions accordingly, $X=\frac{f_\t{crit}}{f}$.

The effect of the collision changes with pooling. The difference between 1 and $X$ collisions is different for different collisional outcomes:

\begin{itemize}
 \item \textit{Bouncing:} Everything as usual but $\delta E\longrightarrow X\cdot\delta E$.
 \item \textit{Coagulation:} Same as for bouncing when it comes to the energy, but the mass and other physical properties of the representative particle changes as $m_i' = m_i + m_k \longrightarrow m_i' = m_i+X\cdot m_k$
 \item \textit{Fragmentation:} For fragmentation we have two cases:
 \begin{enumerate}[label=\arabic*)]
  \item Complete fragmentation of the representative particle with $\leq X$ collisions needed.
  \item Partial fragmentation even when adding up $X$ collisions.
 \end{enumerate}

In Case 1) we assume that $Y=E_\t{frag}/E_\t{coll}$ collisions are required to completely fragment particle $i$ and that the rest of the collisions result in bouncing with the fragment. The new representative particle will have monomer properties and the total energy loss will be $\delta E = \delta E_\t{frag} + \delta E_\t{bounce}$, where $\delta E_\t{frag}$ is the number of physical particles in swarm $i$ times the energy required to completely fragment one of them and $\delta E_\t{bounce}$ is the energy lost from $X-Y$ pooled collisions between a representative particle of monomer mass, $m_0$, and a physical particle of mass $m_k$ (times the number of physical particles in the swarm, $N_i$). Case 2 is more likely and the outcome is calculated just like for one collision but where the collision energy has increased, $E_\t{coll}\longrightarrow X\cdot E_\t{coll}$.
\end{itemize}

\section{Results}\label{sec:sims}

With our simulations we are interested in finding the evolution of the cloud properties during the collapse. We explicitly aim to measure the collapse time as a function of different cloud and pebble parameters, mainly the total mass of the cloud. We also follow the evolution of the size distribution of the pebbles. If the cloud is massive, then the relative speeds between the pebbles will be high and the collisions may result in fragmentation (see \fref{fig:outcomeFig}). From \eref{eq:virialEnergy} we find that the virial speed in the initial state is proportional to the size of the planetesimal which will form,

\begin{align}
v_\t{vir,0}\propto \sqrt{\frac{M}{R_0}}\propto M^{1/3}\propto R_\t{solid}\ . \label{eq:v_vir0}
\end{align}

\noindent Here we assumed that the initial cloud radius is equal to the Hill radius at its semi-major axis. The radius $R_\t{solid}$ is the radius of a body with mass $M$ and density equal to the monomer density. \eref{eq:v_vir0} implies that even if a massive cloud initially consists of cm-sized pebbles, these pebbles can be ground down to $\mu$m-sized dust during the collapse. For low-mass clouds the speeds are lower and collisions will only result in bouncing and the primordial pebbles survive the collapse.

\begin{figure*}
 \begin{center}
  \resizebox{9cm}{!}{\includegraphics{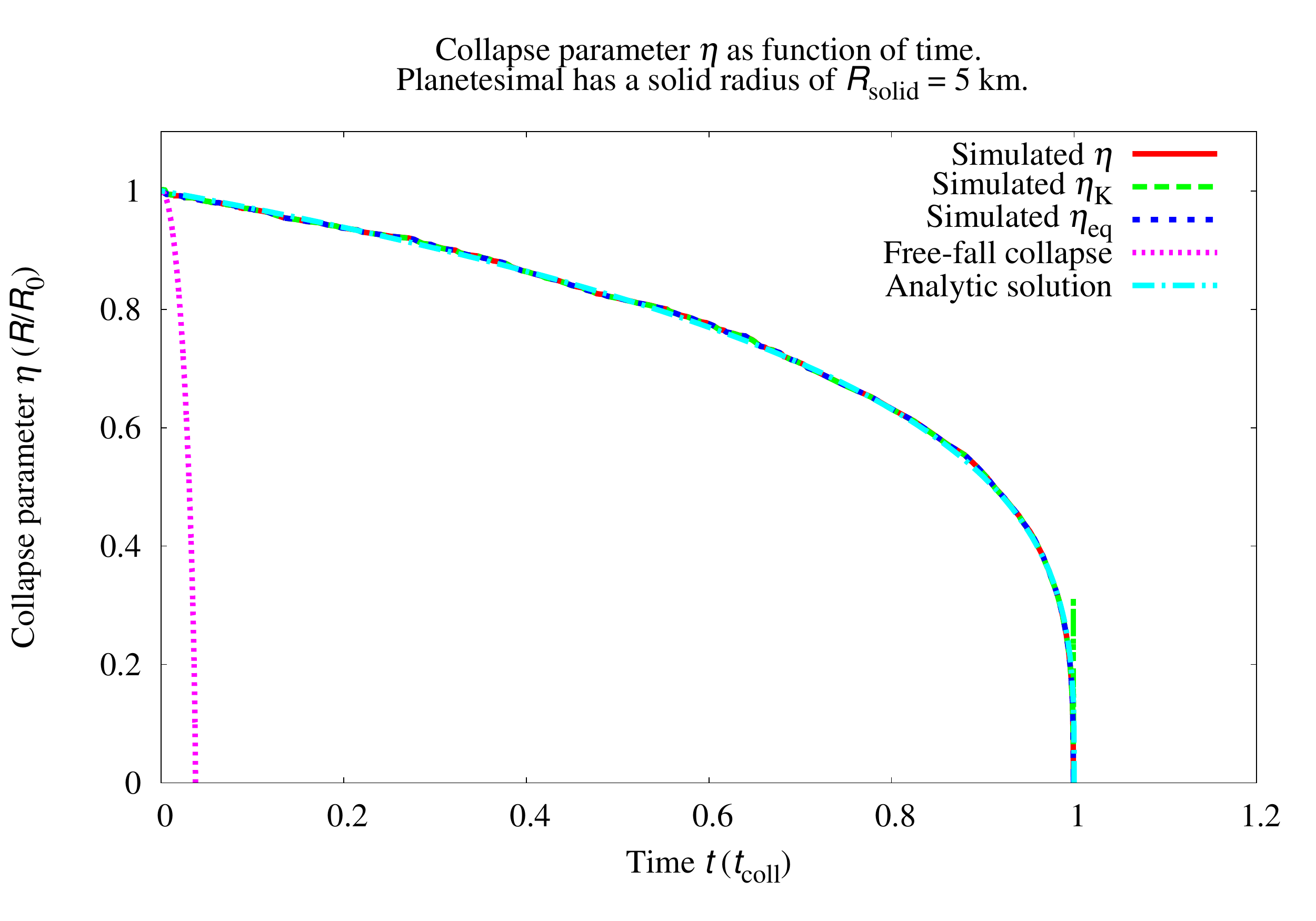}}
  \resizebox{9cm}{!}{\includegraphics{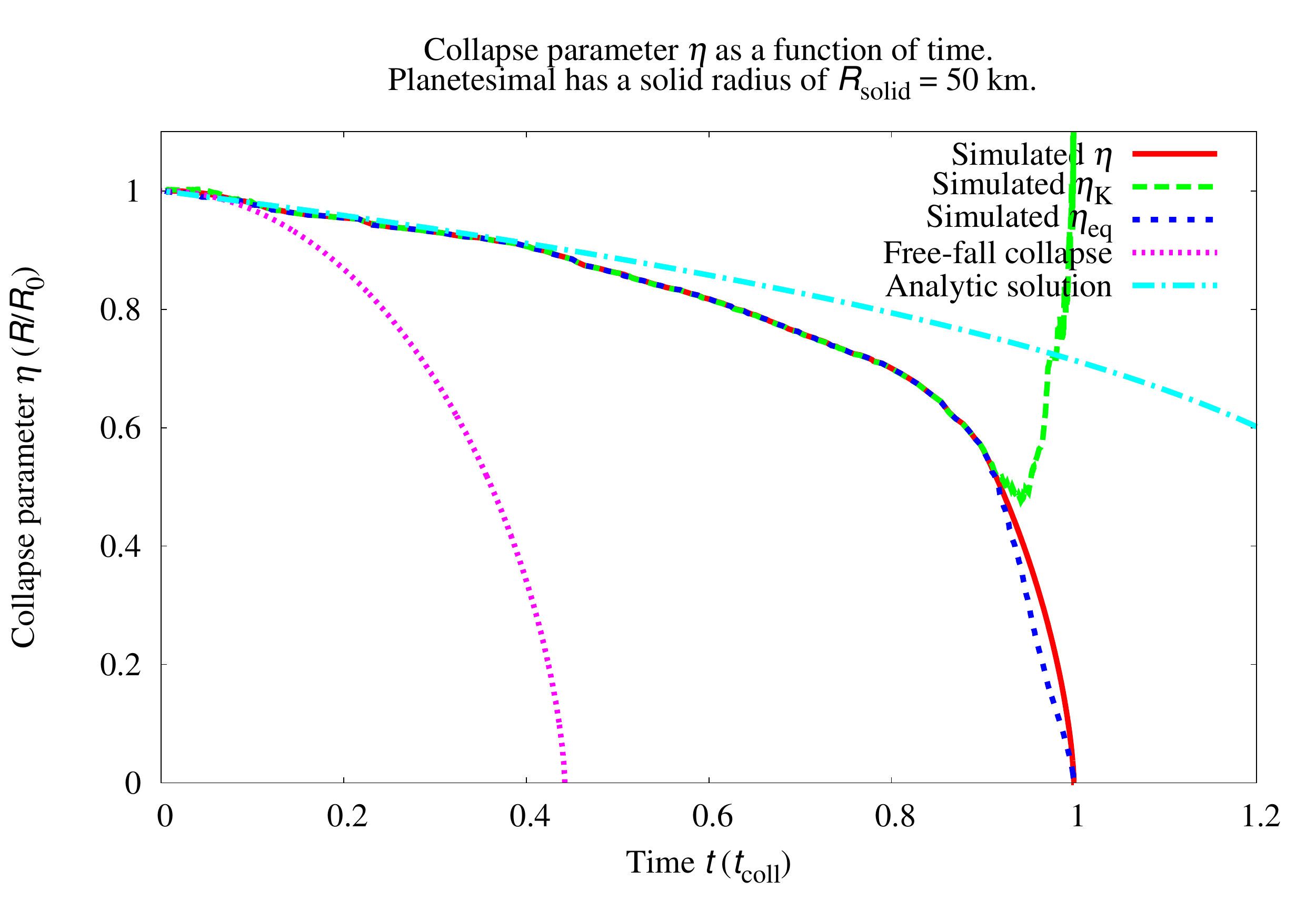}}
  \caption{The total energy, parametrized through $\eta\equiv R(t)/R_0$, as function of time for two clouds with, initially, cm-sized pebbles. The left plot displays the collapse of a cloud with $R_\t{solid}=5$ km and the right plot a cloud with $R_\t{solid}=50$ km. The red lines show the actual size of the cloud, blue lines the equilibrium values ($\eta_\t{eq}\equiv E_0/E$) and green lines the kinetic energy ($\eta_\t{K}\equiv T_0/T$). The pink line shows the free-fall collapse of a cloud and the teal lines the analytic solution from \aref{app:A}. Note that time is measured in units of the collapse time. Therefore the free-fall line is different for the two clouds. In the case of the low-mass cloud the collapse is slow and the only outcome of collisions is bouncing. Therefore it is in virial equilibrium until the very end of the collapse and follows the analytical solution ($\eta=\eta_\t{K}=\eta_\t{eq}$). The massive cloud, on the other hand, has initially higher collision speeds which lead to pebble fragmentation and rapid energy dissipation. At about $\eta=0.5$ the energy dissipation becomes faster than free-fall, so the cloud does not get into virial equilibrium before the next pebble collision (the first term in \eref{eq:deltaEtaT} dominates over the second). This results in a transition to cold collapse ($\eta$ is parallel to the free-fall curve) with sub-virial speeds ($\eta_\t{K}>\eta_\t{eq}$).} \label{fig:collapsePar}
 \end{center}
\end{figure*}

\subsection{Initial conditions and simulation setup}

In our simulations we start out with a self-gravitating cloud of silicate pebbles with a monodisperse size distibution. The pebbles are assumed to be homogeneous spheres (density ${\rho_\t{SiO}}_2$$\sim$$2.5$ g cm$^{-3}$) built up of $\mu$m-sized monomers. The reason for this density is that only the collisional outcome regions for silicates are known in details from laboratory experiments (\fref{fig:outcomeFig}). We assume that the initial radius of the cloud is equal to the Hill radius corresponding to the cloud mass. This means that the density of the cloud, and consequently the free-fall time, is independent of the cloud mass. To be able to use our zero-dimensional model we assume that the cloud is uniform, has no net angular momentum and always strives to get into virial equilibrium. We investigate the planetesimal formation at Kuiper belt distances from the Sun (at orbits with a semi-major axis equal to Pluto's current orbit). Finally we also neglect the influence of any surrounding gas, as the friction timescale is much longer than the collision timescale in those dense clouds.

\subsection{Collapsing pebble clouds} \label{sec:collapse}

The range of cloud sizes we investigate corresponds to solid planetesimals with radii between 1 and 1,000 km, from the smallest planetesimal sizes up to radii comparable to Pluto's. In \fref{fig:tCollFig} we show the results of our simulations of the collapse time, the time at which the cloud reaches monomer density, as a function of planetesimal radius. In the figure we can see that for the simulations with cm-sized pebbles the collapse can be split into three regimes depending on the size of the planetesimal:

\begin{description}
 \item[\textbf{1)} $R_\t{solid}\lesssim 50$ km:] \hfill \\
  The energy dissipation is dominated by bouncing collisions between the primordial pebbles. Collision speeds only reach values high enough for fragmentation to occur in the very end of the collapse. The power-law fit shows us that $t_\t{coll}\propto R_\t{solid}^{-1}$ which is the same dependence as we get from the analytic derivations when we assume that the particle size is constant and bouncing is the only collisional outcome (\aref{app:A}). We find that the collapse time increases linearly with pebble size, which also agrees with the analytic model in \aref{app:A}.
 \item[\textbf{2)} $50\t{ km}\lesssim R_\t{solid}\lesssim 70$ km:] \hfill \\
  In this regime the cloud is massive enough (increased collision speeds) for collisions to result in pebble fragmentation early enough in the collapse to affect the collapse time. From \aref{app:A} we have $t_\t{coll}\propto a/R_\t{solid}$ with constant particle size so the cloud collapses faster if the particles are smaller (increased collision rate). We find that the amount of fragmentation is larger and starts earlier (see \fref{fig:pebble_frac}) the more massive the cloud is (larger initial velocities). This results in a more rapid decrease of the collapse time with increasing cloud mass than in the previous regime.
 \item[\textbf{3)} $R_\t{solid}\gtrsim 70$ km:] \hfill \\
  Collisions are so frequent \erefp{eq:rik} and dissipative \erefp{eq:deltaE} that the collapse is completely limited by the free-fall time of the cloud. Since the clouds initially have the same density, regardless of the cloud mass, above a solid radius of $\sim$70 km all clouds have collapse times equal to the free-fall time.
\end{description}

For the simulations with mm-sized pebbles we only get the first and third regime. The reason for this is that the collapse gets limited by free-fall before it is massive enough for the relative speeds to be high enough for fragmenting collisions. We can also see in \fref{fig:tCollFig} that the collapse time of a cloud in regime 1 is a factor 10 smaller with mm-sized pebbles than with cm-sized pebbles, as expected from the analytic derivations ($t_\t{coll}\propto a/R_\t{solid}$).

The details of a collapse are different depending on the mass of the cloud. \fref{fig:collapsePar} shows the collapse parameter, $\eta$, as a function of time for two different masses. The left plot follows the collapse of a cloud in regime 1 and the right plot a cloud in regime 2. If we would have plotted the collapse of a cloud in regime 3 it would have followed the free-fall curve closely. In the low-mass cloud we can see that the cloud collapses smoothly and the actual size ($\eta$, red curve) follows the equilibrium value ($\eta_\t{eq}$, blue curve). For a cloud of higher mass we run into the problem with virialization and at some point it cannot contract fast enough to keep up with the energy loss in the collisions. At this point $\eta_\t{eq}<\eta$ and $\left|\delta\eta\right|>\left|\delta\eta_\t{max}\right|$ \erefp{eq:deltaEtaMax} and the energy dissipation from this point on is so quick that the cloud free-falls. Hence the $\eta$-curve is parallel to the free-fall curve. Clouds of higher mass in regime 2 have their first fragmenting collision earlier in their collapse and therefore a shorter collapse time.

\begin{figure}
 \begin{center}
  \resizebox{9cm}{!}{\includegraphics{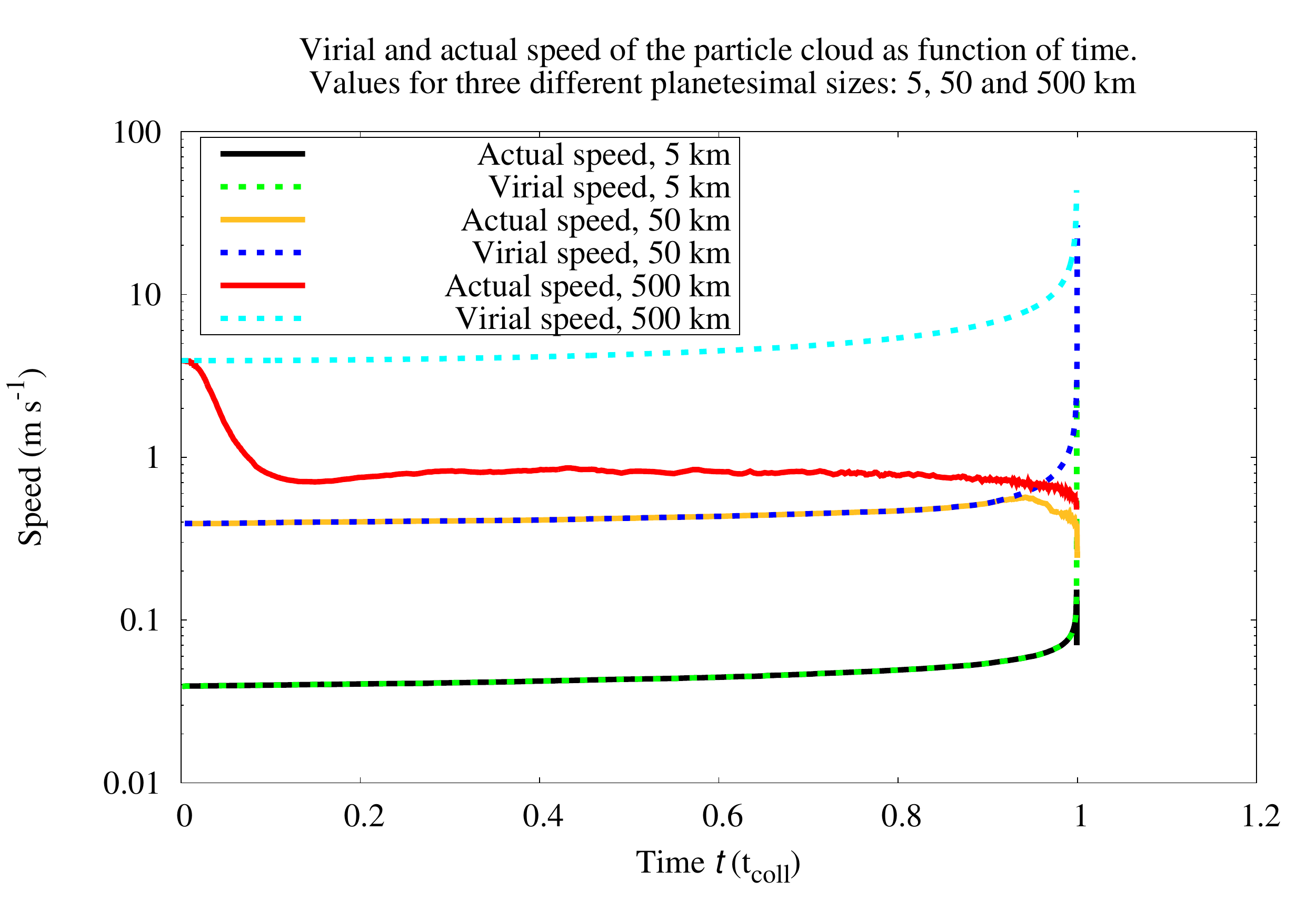}}
  \caption{Virial and actual speeds from $\eta$ and $\eta_\t{K}$ as function of time for three clouds of solid radii of 5 km (green and black), 50 km (blue and gold) and 500 km (cyan and red) with, initially, cm-sized pebbles. The low-mass cloud has a steadily increasing velocity because collisions are infrequent so that virialization can happen for each value of $E$. The massive cloud, on the other hand, dissipates energy too quickly and collapses cold with random pebble speeds much lower than the virial value. The initial virial speed is higher for the more massive planetesimal since the speed is a growing function of total mass (\erefnp{eq:virialEnergy}).} \label{fig:v_vir}
 \end{center}
\end{figure}

\begin{figure}
 \begin{center}
  \resizebox{9cm}{!}{\includegraphics{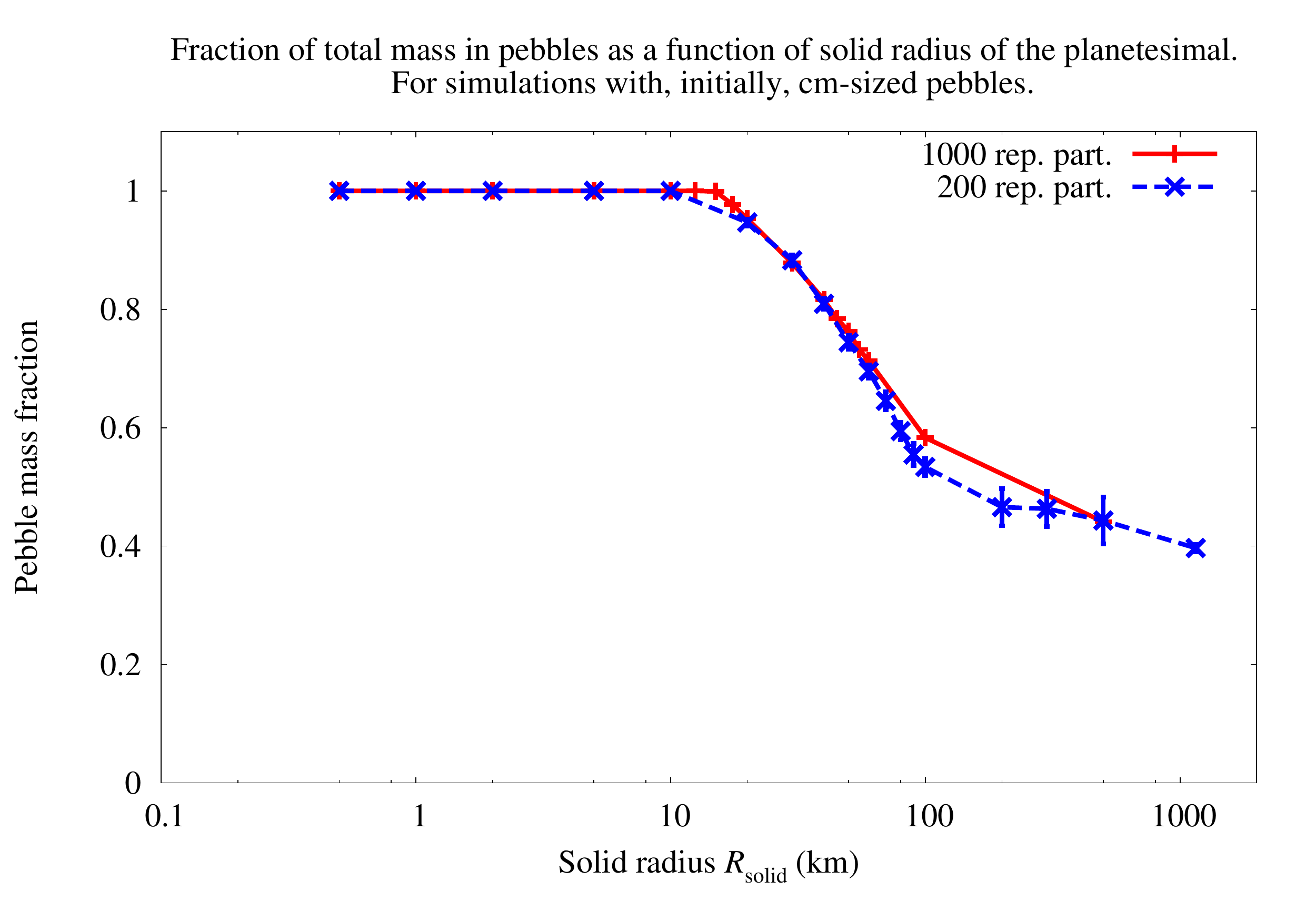}}
  \caption{Pebble mass fraction in the resulting planetesimals for simulations with, initially, cm-sized pebbles, as a function of the planetesimal radius, $R_\t{solid}$. Particles are defined as pebbles if they have radii $a>1$ mm. For more massive planetesimals the collision speeds are larger and the collisions can then result in fragmentation, decreasing the pebble mass fraction. However, massive clouds do not have time to virialize after each collision and the particles will move with sub-virial speeds (see \fref{fig:v_vir}). This causes the amount of fragmenting collisions to decrease as the collapse goes on, ensuring that a fraction of the primordial pebbles survive even in the most massive planetesimals.} \label{fig:pebble_frac}
 \end{center}
\end{figure}

We find in the simulations that both the virialization and the energy dissipation in the collisions are important for the evolution of the kinetic energy. In \fref{fig:v_vir} we plot the evolution of the speed of the particles which we get from \eref{eq:etaK}, assuming that all particles have equal speeds. For the cloud in regime 1 (black curve) the speeds are steadily increasing thanks to the virialization of the cloud except for the very end of the collapse where virialization cannot keep up. For the cloud in regime 2 (yellow curve), on the other hand, the collapse is, as we have already seen, somewhat limited by free-fall and the speeds decrease thanks to the energy dissipation in the collisions (right panel in \fref{fig:collapsePar}). In this case the first term in \eref{eq:deltaEtaT} dominates over the second term. This effect is greater the more massive the cloud is. The particles in the cloud in regime 3 (black curve) move with sub-virial speeds very early in the collapse and the collision speeds reach values low enough for coagulation. We also see that the initial virial speed scales linearly with planetesimal radius which originates from the fact that the clouds initially have a radius equal to their Hill radius and that they are in virial equilibrium so $v_\t{vir,0}\propto R_0$ \erefp{eq:virialEnergy}.

Finally we are also interested in the sizes of the particles in the resulting planetesimal. If the planetesimal is built up by cm-sized pebbles it will be porous and have both low density and low internal strength. Such an object could be tidally broken up like the Shoemaker-Levy 9 comet \citep{asphaug96}. If, on the other hand, the planetesimal is a mixture of pebbles and dust it can be packed very compactly and have a high density. In the case of low-mass clouds they collapse without any fragmentation and therefore become pebble piles with low density. This is in agreement with both the observed mass-density relation of Kuiper belt objects \citep{brown13}, low-mass Kuiper belt objects have low density, and the model of comets as pebble piles with low internal strength \citep[e.g.][]{skorov12,blum14}. The density of planetesimals can, however, evolve as time goes on after the collapse through other physical mechanisms. Self-gravity can cause static compression and an increase in the density of the planetesimals \citep[e.g.][]{kataoka13}. Another mechanism that could affect both the internal structure and pebble mass fraction of the planetesimals is radioactive heating. Short-lived radioactive isotopes such as $^{26}$Al could supply enough heat for differentiation to occur. 

For a high-mass cloud fragmentation of pebbles is occurring during the collapse and the planetesimal can be more tightly packed and have a higher density, again consistent with the higher densities of larger Kuiper belt objects. In \fref{fig:pebble_frac} we plot, for simulations with initially cm-sized pebbles, the pebble mass fraction (particles with radii $a>1$ mm) in the resulting planetesimals as a function of the size of the planetesimals. First of all we see that planetesimals with radii $R_\t{solid}\lesssim 20$ km are completely built up by pebbles. Planetesimals slightly larger, $20\t{ km}\lesssim R_\t{solid}\lesssim 50\t{ km}$, are still in collapse regime 1 and follow the analytic approximations relatively well. The reason for this is the fact that the first fragmenting collision occurs late in the collapse and energy dissipation is dominated by bouncing collisions between cm-sized pebbles. Next we see that the pebble mass fraction decreases with increasing planetesimal size. However, even for the most massive planetesimals we still have a large fraction of the mass in pebbles. The reason that not all pebbles are ground down to dust is the fact that the particles move with sub-virial speeds in the end of the collapse (see \fref{fig:v_vir}) and the collision speeds are not high enough for fragmentation to occur. In the end of the simulations most collisions result in bouncing or coagulation.

\begin{figure}
 \begin{center}
  \resizebox{9cm}{!}{\includegraphics{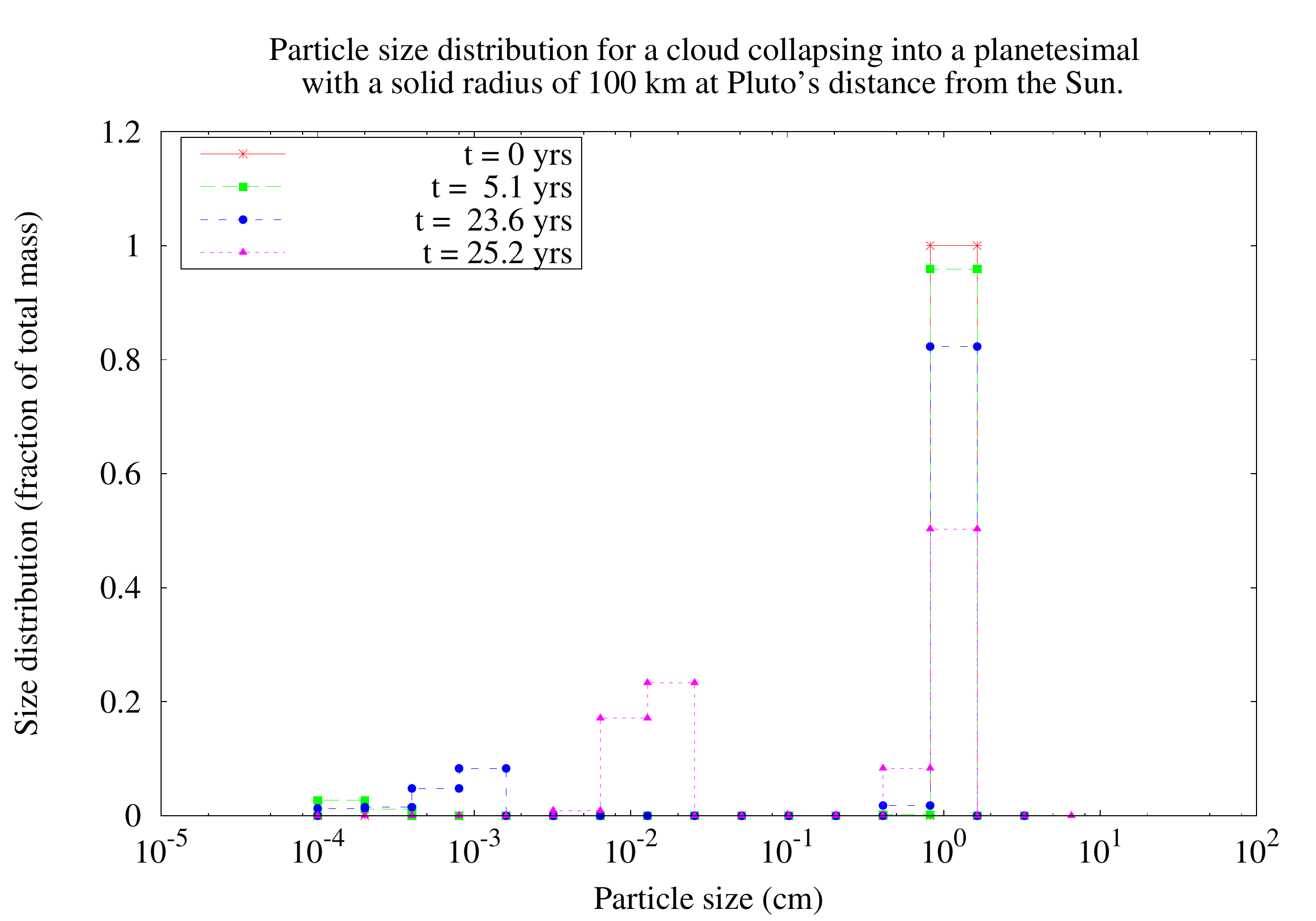}}
  \caption{The particle size distribution in a collapsing pebble cloud with solid radius $R_\t{solid}=100$ km at four different times. The collision speeds are initially high enough for fragmentation to occur. Later in the collapse the particles move with sub-virial speeds (see \fref{fig:v_vir}) and the amount of fragmentation decreases. The low collision speed in the cold collapse allow the dust to coagulate and grow to sizes of $\sim$0.1 mm. One thing we also see in the figure is that, at the end of the collapse, a small fraction of the primordial pebbles have been subject to erosion through collisions with dust and decreased in size.} \label{fig:size_distr}
 \end{center}
\end{figure}

The evolution of the particle size distribution in a simulation of a cloud collapsing into a 100 km-sized planetesimal is shown in \fref{fig:size_distr} and can broadly be summarized as follows. The cloud starts with a monodisperse size distribution of cm-sized pebbles. As times goes on, pebbles collide with high enough velocities to fragment down to $\mu$m-sized dust. The sub-virial collision speeds in the cold collapse phase ensures that a fraction of the pebbles survive. Collisions involving the dust can result in sticking and the dust coagulate and grow up to $\sim$mm-sized pebbles again. In the end there is a roughly bimodal size distribution consisting of some primordial pebbles have survived the collapse and the growing dust that originally was ground down from the primordial pebbles. 

\section{Conclusions and Discussion}\label{sec:conc}

In this paper we have investigated the internal evolution of pebble clouds formed in protoplanetary disks by the streaming instability, although the results can be applied to other planetesimal formation models relying on particle concentration and gravitational collapse as well \citep[e.g.][]{klahr03, cuzzi08, dittrich13}. We have developed a numerical model to study the effect of dissipative collisions between pebbles leading to gravitational collapse of the cloud. The main results of our simulations can be summarized as follows.

The collapse times of the pebble clouds in our model are short. We assume that all clouds initially have a size equal to their Hill radius, giving a free-fall time of $t_\t{ff}\sim 25.1$ years at $\sim$40 AU. The collapse time still varies depending on the cloud properties. Massive clouds ($R_\t{solid}\gtrsim70$ km) collapse, roughly, on the free-fall time of the cloud and even for low-mass clouds ($R_\t{solid}$$\sim$km) the collapse time is only a few tens of orbits. The size of the pebbles affects the collapse time, with more rapid collapse for smaller pebbles (due to their larger collision surface). These collapse timescales can be compared to the formation times of the pebble clouds which are just a few ten years \citep{johansen09}.

The clouds do not get into virial equilibrium immediately after a dissipative collision but require some time to fall to their new (smaller) desired size. The result is that the particles in high-mass clouds will move at sub-virial speeds, reducing their collision speeds significantly and preventing widespread fragmentation. For a low-mass cloud this does not occur during most of the collapse, since the collision rates are low and the cloud has time to virialize between two collisions. It is only at the end of the collapse, when the density of the cloud is very high, that the particles in low-mass clouds dissipate energy quickly enough that they will move with sub-virial speeds. A high-mass cloud, on the other hand, enters the cold collapse phase already in the beginning of the contraction. The result is that massive clouds collapse coldly and after an initial burst of pebble fragmentation the amount of fragmenting collisions decreases and the dust can start to grow to reform larger and larger aggregates (up to $\sim$mm).

In the collapse of low-mass clouds ($R_\t{solid}\lesssim 20$ km) the collision speeds never reach the fragmentation limit and they will be completely made up of the primordial pebbles. In more massive clouds ($R_\t{solid}\gtrsim 20$ km) the collision speeds are, at some point in the collapse, large enough for fragmentation to occur in the pebble collisions. In all planetesimals some of the primordial pebbles survive. For the mid-size planetesimals ($20\t{ km}\lesssim R_\t{solid}\lesssim 50\t{ km}$) the first fragmenting collision occurs late in the collapse and not all pebbles have time to be ground down. In the most massive clouds even the first collisions are fragmenting but slow virialization of the cloud causes the collision speeds later in the collapse to be smaller and a significant fraction of the pebbles survive. This formation process causes an observable difference between a low- and a high-mass planetesimal, namely the bulk density. Low-mass planetesimals can be thought of as pebble piles and they will be very loosely packed and have a low density. This will also make them easily tidally disrupted observed in Jupiter's tidal disruption of the comet Shoemaker-Levy 9. More massive planetesimals, containing both dust and pebbles, are able to become very tightly packed and therefore achieve a high density. This relationship has been suggested for objects in the Kuiper belt \citep{brown13} but we stress that there are more parameters that affect the density, such as composition, differentiation through radioactive heating and collisional impacts.

We use the results from laboratory experiments of collisions between silicate aggregates even if Kuiper belt objects are mainly composed of ice. One difference is that ice particles could be very fluffy and survive collisions at higher speeds compared to silicates \citep[e.g.][]{wada09,wada13}. However, even in our most massive clouds the collision speeds do not reach values over $\sim$10 m s$^{-1}$ and the speed decreases rapidly as the collapse progresses (red line in \fref{fig:v_vir}). The collapse times, on the other hand, could very well be affected by our choice of material. In \aref{app:A} we can see that for a given cloud mass and pebble size the collapse time decreases for ice since the material density is lower (the collision rates increases) but the main conclusion that pebbles survive the collapse would not change.

The next step in our investigation will be to increase the dimensionality of the simulations. Our current numerical model is zero-dimensional in the sense that we keep track only of the cloud size and collisions are considered via a Monte Carlo model. The 0-D approach is not completely physically correct since the pebble cloud should initially have some rotation which will prevent the direct collapse. However, we expect that the main findings of the paper will be true also in a 3-D model with rotation, since gravitational instabilities in the rotating pebble disk should lead to angular momentum transport and the excitation of relative speeds which can drive a collapse similar to the 0-D model. Hydrodynamical 2-D and 3-D models will be presented in a subsequent paper.

\begin{acknowledgements}
KWJ and AJ were supported by the European Research Council under ERC Starting Grant agreement 278675-PEBBLE2PLANET. AJ was also supported by the Swedish Research Council (grant 2010-3710) and by the Knut and Alice Wallenberg Foundation. KWJ and AJ acknowledges the support from the Royal Physiographic Society in Lund for grants to purchase computer hardware to run the simulations on. We are grateful for stimulating discussions with Jürgen Blum, Hubert Klahr, Tristen Hayfield, Patrick Glaschke and Thomas Henning. We also thank the referee Hiroshi Kobayashi for many insightful comments that helped improve the manuscript.
\end{acknowledgements}

\bibliographystyle{aa} 
\bibliography{myRef} 

\begin{appendix}
\section{Analytic derivation of the collapse of a particle cloud}\label{app:A}

As mentioned in the main text, it is possible to derive the evolution of a collapsing cloud analytically under some assumptions. First of all we assume a uniform density and no net rotation so that we can make the model 0-dimensional. Next we assume that we have equal particle sizes, equal speeds ($\bar{v}$) and that the only outcome of a collision is bouncing. The last point is only valid for low-mass clouds which exhibit low collision speeds. A final assumption is that the cloud virializes immediately after a collision. This, we will later see, makes the relative speeds too high and the collapse time too small for clouds that are not in regime 1 (see Section \ref{sec:collapse}).

We estimate the evolution of the energy of the cloud with the equation

\begin{align}
 \frac{\d E}{\d t}=\delta E\times r_\t{coll}\ , \label{eq:EDot1}
\end{align}

\noindent where $\delta E$ is the change in energy per collision and $r_\t{coll}$ is the collision rate (number of collisions per unit time). The energy change per collision is (from \erefnp{eq:deltaE})

\begin{align}
 \delta E &=-\frac{1}{8}m\left(1-C_\t{R}^2\right)\Delta v^2=-A\Delta v^2\ ,  &A\equiv\frac{1}{8}m\left(1-C_\t{R}^2\right) \label{eq:dE(v)}
\end{align}

\noindent where $m$ is the mass of each individual particle. The collisions are not necessarily head-on which makes the effective collision speed smaller and reduces the average energy dissipated with a factor $1/2$. To get the collision rate of the system we look at the rate with which two equal-sized particles in a volume $V$ collide:

\begin{align}
r_\t{ik}&=\frac{\sigma_\t{ik}\Delta v}{V}=\frac{3}{4\pi R^3}4\pi a^2\Delta v=3a^2\frac{\Delta v}{R^3}\ , \label{eq:rikApp}
\end{align}

\noindent where $\sigma_\t{ik}$ is the geometrical cross-section of particle $i$ and $k$, $a$ is the particle radius and $R$ is the radius of the cloud. The total rate is the sum of the rates of all possible collisions (a particle cannot collide with itself and we cannot count both $r_\t{ik}$ and $r_\t{ki}$ as it is the same collision), and setting $r_\t{ik}$ constant for all collision pairs, we get

\begin{align}
r_\t{coll}&=\sum_{i=1}^{N_\t{p}-1}\sum_{k=i+1}^{N_\t{p}}r_{ik}=r_{ik}\sum_{i=1}^{N_\t{p}-1}\sum_{k=i+1}^{N_\t{p}}=r_{ik}\sum_{i=1}^{N_\t{p}-1}(N_\t{p}-i) \nonumber \\
        &=r_\t{ik}\left(N_\t{p}(N_\t{p}-1)-\sum_{i=1}^{N_\t{p}-1}i\right)=\frac{N_\t{p}(N_\t{p}-1)}{2}r_\t{ik} \nonumber \\
        &=\frac{3}{2}N_\t{p}(N_\t{p}-1)a^2\frac{\Delta v}{R^3}\approx\frac{3}{2}N_\t{p}^2a^2\frac{\Delta v}{R^3}\ . \label{eq:rColl}
\end{align}

\noindent The last step is valid since the number of particles $N_\t{p}\gg 1$. We then write the rate as

\begin{align}
r_\t{coll}&=B\frac{\Delta v}{R^3}\ ,  &B\equiv \frac{3}{2}\frac{M^2}{m^2}a^2=\frac{3}{2}\frac{R_\t{solid}^6}{a^4}\ . \label{eq:rTot}
\end{align}

\noindent Here $M$ is the total mass of the cloud and $R_\t{solid}$ is the radius of a planetesimal with mass $M$ and density equal to the internal density of the particles. Now \eref{eq:EDot1} becomes

\begin{align}
 \frac{\d E}{\d t}=-AB\left(\frac{\Delta v}{R}\right)^3\ . \label{eq:EDot2}
\end{align}

\noindent To get $\Delta v$ and $R$ as functions of the total energy we use \eref{eq:virialEnergy}

\begin{align}
 -E&=\frac{M}{2}\bar{v}^2=C\Delta v^2\ ,&C&=\frac{M}{4}\ , \label{eq:E(v)}\\
 -E&=\frac{D}{R}\ ,&D&=\frac{3}{10}GM^2 \label{eq:E(R)}\ ,
\end{align}

\noindent where the change from $\bar{v}$ to $\Delta v$ comes from the fact that the relative speed is on average $\sqrt{2}$ times larger than the individual speeds. Inserting \erefCon{eq:E(v)}{eq:E(R)} into \eref{eq:EDot2} yields

\begin{align}
 \frac{\d E}{\d t}&=-\underbrace{ABC^{-3/2}D^{-3}}_{K'}(-E)^{9/2}\nonumber \\
                  &\Longrightarrow (-E)^{-9/2}\d(-E)=K'\d t \nonumber \\
                  &\Longrightarrow -\frac{2}{7}(-E)^{-7/2}=K't-Q' \nonumber \\
                  &\Longrightarrow -E=\frac{1}{(Q-Kt)^{2/7}}\ . \label{eq:E(t)}
\end{align}

\noindent Here $K=3.5K'$ is a measure of the rate with which the energy changes and $Q=3.5Q'$ is a measure of the initial energy. To get the value of the parameter $Q$ we look at the initial energy, $E_0=E(t=0)$, and using \eref{eq:virialEnergy} we obtain

\begin{align}
\left. 
\begin{array}{l}
 -E_0=Q^{-2/7} \\
 -E_0=-\frac{U_0}{2}=\frac{3}{10}\frac{GM^2}{R_0}
\end{array}\right\}\Longrightarrow Q=\frac{2^{7/2}\cdot 5^{7/2}}{3^{7/2}}\frac{R_0^{7/2}}{G^{7/2}M^7}\ , \label{eq:Q}
\end{align}

\noindent where $R_0$ is the initial radius of the cloud. A quick look at \eref{eq:E(t)} shows us that if $t=t_\t{coll}\equiv Q/K$ then $-E\longrightarrow\infty$, i.e. a complete collapse ($R\longrightarrow 0$ and $\bar{v}\longrightarrow\infty$). Finally we assume that the initial radius of the cloud is equal to the Hill radius at the semi-major axis of the cloud

\begin{align}
 R_0=R_\t{Hill}(d)=\left(\frac{M}{3M_\odot}\right)^{1/3}d\ , \label{eq:RHill}
\end{align}

where $d$ is the semi-major axis of the cloud's orbit. Now we rewrite \eref{eq:E(t)} as 

\begin{align}
 E=\frac{E_0}{\left(1-\frac{t}{t_\t{coll}}\right)^{2/7}}\ .
\end{align}

By using Eqs. \eqref{eq:dE(v)}, \eqref{eq:rTot}, \eqref{eq:E(v)}-\eqref{eq:RHill}, assuming silicate pebbles and placing the cloud at Pluto's distance from the Sun we can find an expression for the collapse time

\begin{align}
 t_\t{coll}=\frac{Q}{K}=4.1\t{ kyr}\left(\frac{R_\t{solid}}{1\t{ km}}\right)^{-1}\left(\frac{a}{1\t{ cm}}\right)\left(1-C_R^2\right)^{-1}\ . \label{eq:tCollApprox}
\end{align}

\end{appendix}

\end{document}